\documentclass{aa} 
\usepackage{graphicx}
\usepackage[sectionbib,authoryear]{natbib} 
\usepackage{times}
\usepackage{graphicx} 

\bibpunct{(}{)}{;}{a}{}{,} 

\newcommand{\lsim}{\,\raisebox{0.2em}{$<$}\!\!\!\!\!
\raisebox{-0.25em}{$\sim$}\,}
\newcommand{\gr}{$\gamma$-ray \,}
\newcommand{\grs}{$\gamma$-rays \,}
\begin{document}
  \title{Cosmic ray acceleration
    parameters from multi-wavelength observations. The case of SN~1006}

\titlerunning{CR acceleration parameters of SN~1006}

  \author{E.G.Berezhko
          \inst{1}
          \and
           L.T.Ksenofontov
          \inst{1}
          \and
          H.J.V\"olk
          \inst{2}}

   \offprints{H.J.V\"olk}

   \institute{Yu.G.Shafer Institute of Cosmophysical Research and Aeronomy,
                     31 Lenin Ave., 677980 Yakutsk, Russia\\
              \email{berezhko@ikfia.ysn.ru}
              \email{ksenofon@ikfia.ysn.ru}
         \and
             Max-Planck-Institut f\"ur Kernphysik,
                Postfach 103980, D-69029 Heidelberg, Germany\\
             \email{Heinrich.Voelk@mpi-hd.mpg.de}
             }

   \date{Received month day, year; accepted month day, year}

   \abstract{}
{The properties of the Galactic supernova remnant (SNR) SN~1006
     are theoretically reanalysed.}  
{Nonlinear kinetic theory is used to
     determine the acceleration efficiency of cosmic rays (CRs) in the
     supernova remnant SN~1006. The known range of astronomical parameters and
     the existing measurements of nonthermal emission are examined in order to
     define the values of the relevant physical parameters which determine the
     CR acceleration efficiency.}
   {It is shown that the parameter values -- proton injection rate, electron to
     proton ratio and downstream magnetic field strength --are determined with
     the appropriate accuracy. In particular also the observed azimuthal
     variations in the \gr morphology agree with the theoretical
     expectation. These parameter values, together with the reduction of the \gr
     flux relative to a spherically symmetric acceleration geometry, allow a
     good fit to the existing data, including the recently detected TeV
     emission by H.E.S.S.}
   {SN~1006 represents the first example where a high efficiency of nuclear CR
     production, required for the Galactic CR sources, is consistently
     established.}
 

\maketitle

%

\section{Introduction}

Cosmic rays (CRs) below the energy $~10^{17}$~eV are believed to be accelerated
in the shell-type supernova remnants (SNRs) of our Galaxy \citep[e.g.][]{bv07}
by means of the diffusive shock acceleration process. However, this
proposition has still only a limited observational/theoretical basis. In
  addition SNRs may not only act individually but possibly also more
  collectively in the environment of interacting stellar winds and supernova
  explosions in OB associations \citep[e.g.][]{parizot04,binns07}. To ensure
that Galactic SNRs are indeed efficient sources of the
Galactic CRs to begin with a number of SNRs
is required with clearly determined astronomical parameters, like the
type of supernova explosion, the SNR age, the distance, and the properties of
the circumstellar medium.  Applying to such SNRs an appropriate model which
consistently describes the dynamics of the SNR, one can then predict the
properties of the accelerated particles and the dynamical and
radiative effects which they produce, like the shock modification and the
multi-wavelength nonthermal emission.

The success of the theoretical model can be judged by
comparison with the experimentally determined overall broad band spectrum and
with the morphological characteristics of the SNR, like its filament structures
and the general radial and azimuthal variations of the emission pattern, as
well as the internal dynamics characterized by the contact discontinuity
between ejected and swept-up mass. From the point of view of Galactic CR
origin, the key quantity is the efficiency of CR production.

In practice such a program is most often hampered by the
limited amount and detail of relevant observations.  First of all, the
astronomical parameters of SNRs are as a rule poorly known. Even though the SNR
age is known for several historical SNRs, the distance is usually quite
uncertain.  In this regard SN~1006 is an exception: the distance was determined
using optical measurements with relatively high precision \citep{winkler03}.

The second problem is that even the presently most advanced nonlinear kinetic
theory of CR acceleration in SNRs \citep{bek96,bv97} \footnote{These papers
  contain the basic equations used, in the form of a transport equation for
  energetic charged particles in a shock environment and its coupling to the
  gas dynamical equations for the thermal plasma. Considering seperately the
  energetic electron component, a corresponding transport equation is added
  that includes radiative energy losses \citep{bkv02}. Also the particles'
  diffusion properties in the form of the Bohm limit, wherever nuclear
  particles are effectively injected into strong shocks, as well as the assumed
  heating rate of the thermal plasma by wave dissipation are detailed in these
  papers.}  contains physical parameters which can not yet be theoretically
calculated with the necessary precision \citep[see][for a
review]{voelk04}. This concerns the magnitude and the spatial distribution of
the injection rates of ions and of electrons into the diffusive shock
acceleration process as well as the extent of magnetic field amplification in
this process.  Fortunately, the values of these parameters can be inferred from
the observed radio and X-ray synchrotron spectra if they are measured in
sufficient detail \citep[see][for reviews]{ber05,ber08}. In such a case this
theory provides a consistent if still approximate model of both SNR and CR
dynamics, and of the properties of the emission produced by the accelerated
particles. In particular, the theory {\it predicts} the high-energy \gr
spectrum.

  Up to now SN~1006 is the only example, for which all astronomical parameters
  are quite well known \citep[e.g.][]{cassam08}. In addition the
  nonthermal emission in the radio and X-ray bands has by now been measured
  quite accurately with Chandra \citep{allen04,allen08} and Suzaku
  \citep{bamba08}. Beyond that, the VHE emission of SN~1006 has been recently
  detected with H.E.S.S., both regarding its flux and its morphology
  \citep{naumann09}. This makes SN~1006 uniquely
  suitable for theoretical study and for a detailed comparison with the
  experimental data.

  We have already applied this nonlinear theory to the case of SN~1006 before
  \citep{bkv02,bkv03,vbk03,kbv05,vbk05,vkb08}. In the following a brief review
  of this previous work is given. The corresponding model assumes spherical
  symmetry and a quasi-parallel shock. It was, however, argued that efficient
  nuclear particle injection/acceleration should arise only within two polar
  cap regions, where the SN shock is quasi-parallel with respect to the ambient
  ISM magnetic field which in turn is approximately perpendicular to the line
  of sight \citep{vbk03}. The calculated size of the efficient nuclear CR
  production regions, which amounts to about 20\% of the shock surface,
  corresponds very well to the observed sizes of the bright X-ray synchrotron
  emission regions. Over the remaining 80\% of the surface the shock is
  quasi-perpendicular and this leads to a depression of the injection of
  nuclear particles into the diffusive shock acceleration process. This means
  that efficient nuclear CR production -- that occurs in a spherically
  symmetric model uniformly across the whole shock surface -- in the actual SNR
  takes place only within the polar regions. It is therefore consistent with a
  correction for the spherically symmetric solution for the nuclear CR
  production and the associated \gr emission by a renormalization factor
  $f_\mathrm{re} \approx 0.2$ (for details, see the Appendix; to emphasize the
  three-dimensional character of the configuration a description in terms of
  magnetic flux tubes is given.). In the first study \citep{bkv02} it was
  demonstrated that the observed properties of the nonthermal emission can be
  consistently understood assuming a considerably amplified magnetic field
  $B_d\approx 120$~$\mu$G within the nuclear injection/acceleration
  region. Such a strong magnetic field is consistent with theoretical
  expectation and was confirmed by the Chandra detection \citep{bamba03,long03}
  of filament structures in the nonthermal X-ray emission of the rims of the
  SNR shell \citep[see][for an interpretation]{bkv03}. A filamentary structure
  of this character is indeed expected in young SNRs, where the magnetic field
  is strongly amplified, as the result of the strong synchrotron losses of the
  X-ray emitting CR electrons. In fact, since there exists generally a simple
  relation between the filament thickness and the interior magnetic field
  strength \citep{bv04a}, the measurement of the filament thickness represents
  a new and independent method for the determination of the magnetic field
  strength inside young SNRs. It is important to note that the interior
  magnetic field strength, determined in such a way for all known Galactic
  SNRs, agrees very well with its value derived from a fit of the shape of the
  overall synchrotron spectra, wherever both methods can be applied
  \citep{vbk05}.  It was also shown \citep{bkv02} that nonlinear kinetic theory
  of CR acceleration is consistent with the TeV-emission detected by CANGAROO
  in 1998 for a value $N_\mathrm{H}=0.3$~cm$^{-3}$ of the ambient interstellar
  medium ISM density from the range $0.05 \leq N_\mathrm{H} \leq 0.3$~cm$^{-3}$
  existing in the literature. However, SN~1006 was not confirmed as a TeV
  source by the H.E.S.S. experiment in a total of 18.2h (in 2003) and 6.3h (in
  2004) ``livetime'' of ON source observations with SN~1006 in the field of
  view \citep{aharonian05}. The H.E.S.S. upper limit was roughly one order of
  magnitude lower than the published CANGAROO flux.  In a subsequent paper by
  the present authors \citep{kbv05} it was demonstrated that this
  H.E.S.S. upper limit does not invalidate the theoretical picture on which the
  previous calculation of the \gr emission spectrum had been based. As it was
  shown, it is rather the value of the external astronomical parameter
  $N_\mathrm{H}$ that strongly influences the hadronic \gr flux: the hadronic
  \gr flux is very sensitive to the ambient gas density $N_{\mathrm H}$ and the
  H.E.S.S. upper limit requires $N_{\mathrm H} < 0.1$~cm$^{-3}$. It is
  important to realize that only a rather strongly amplified magnetic field,
  produced nonlinearly by efficiently accelerated nuclear CRs, is compatible
  with such a low \gr flux which is therefore expected to be predominantly of
  hadronic origin. Otherwise the overproduction of accelerated electrons would
  leed to an unacceptably high inverse Compton (IC) \gr emission
  \citep{bkv02}. This requires a strong component of accelerated nuclear
  particles whose energy density substantially exceeds that of the synchrotron
  electrons. The opposite simplest IC scenario, which is based on the
  assumption that the entire nonthermal SNR emission is produced by accelerated
  electrons without any substantial nuclear CR production and magnetic field
  amplification, substantially overpredicts the existing \gr upper limits for
  all known type Ia Galactic SNRs \citep{vkb08}.

  Since TeV \gr emission of SN~1006 was recently detected \citep{naumann09} it
  is worthwhile to reconsider SN~1006 on the basis of this nonlinear
  theory. Compared with the previous consideration \citep{kbv05} this makes it
  possible to determine the ambient ISM density within narrow limits.  In
  addition, we shall demonstrate that the values of the relevant physical
  parameters -- proton injection rate, electron to proton ratio and downstream
  magnetic field strength -- are determined for SN~1006 with appropriate
  accuracy from the measured synchrotron spectrum. The \gr morphology agrees
  with the theoretical expectations regarding the morphology of ion injection
  and the corresponding morphology of magnetic field amplification
  \citep{vbk03}. It is therefore indeed consistent with a correction for the
  total nuclear particle pressure and the total \gr flux obtained in the
  spherically symmetric solution by a renormalization factor $f_\mathrm{re}
  \approx 0.2$.  The correction implies the full value of particle pressure and
  \gr flux, as calculated in the spherically symmetric model, in the 
    quasi-parallel part of the magnetic flux tubes that
    thread the polar cap regions, but a value of zero elsewhere. The above
  parameter values plus the mentioned morphological aspects allow a physical
  explanation of all existing multi-wavelength data, including the
  H.E.S.S. measurement of the TeV spectrum. There is possibly one exception to
  this conclusion. It concerns the radius of the contact discontinuity between
  ejected and swept-up mass relative to the radius of the SNR blast wave,
  cf. recent data discussed by \citet{cassam08} which indicate a large value --
  apparently quite close to unity -- of this ratio. However another very recent
  analysis \citep{miceli09} claims values $\sim 0.9$ for this ratio which are
  rather in line with quantitative expectations for a CR-modified shock in an
  object of comparable evolutionary phase like Tycho's SNR
  \citep{warren05,cassam07,vbk08}.  Future work will have to assess these
  differences.

\section{Results and discussion}
Since SN~1006 is a type Ia supernova it presumably expands into a uniform ISM,
ejecting roughly a Chandrasekhar mass $M_\mathrm{ej}=1.4 M_{\odot}$. Since the
gas density indeed varies only mildly across the SNR \citep {acero07}, it
appears reasonable to assume also the circumstellar magnetic field to be
uniform. The ISM mass density $\rho_0=1.4m_\mathrm{p}N_\mathrm{H}$, which is
usually characterized by the hydrogen number density $N_\mathrm{H}$, is an
important parameter which strongly influences the expected SNR dynamics and
nonthermal emission.

As in our earlier study \citep{kbv05} we  solve here the
  coupled set of nonlinear equations. mentioned before, in order to find the
optimum set of physical parameters of SN~1006 which gives a consistent
description of the observed overall dynamics and of the nonthermal emission
together with its morphology. The theory includes all the important physical
factors which influence CR acceleration and SNR dynamics: shock modification by
CR backreaction, MHD wave damping within the shock transition, a consistently
determined CR spectrum, and the spatial distributions in each evolutionary
phase. In addition it includes synchrotron losses of CR electrons and a
determination of all nonthermal emission processes, produced in SNRs by the
accelerated CRs.  It had also been shown that the values of 
the key parameters (proton injection rate $\eta$, electron to proton ratio
$K_\mathrm{ep}$ and interior (downstream) magnetic field strength
$B_\mathrm{d}$) which can not be predicted theoretically with the required
accuracy, can be determined from a fit of the observed synchrotron emission
data. It is of basic importance here that the parameter values for SN~1006,
determined in this way, were very well confirmed by the Chandra measurements of
the fine structure of the nonthermal X-ray emission \citep{bamba03,long03}, as
analyzed by \citet{bkv03}. 

Compared with the previous study \citep{kbv05} the analysis includes the
  most accurate X-ray data from Chandra \citep{allen04} and Suzaku
gv   \citep{bamba08} that make it possible to quite precisely determine the
acceptable range of these parameter values. Also a time-dependent,
  amplified upstream magnetic field
\begin{equation}
B_0(t)=B_0(t_\mathrm{sn})\sqrt{P_\mathrm{c}(t)/P_\mathrm{c}(t_\mathrm{sn})}
\end{equation}
is used -- in the spherically symmetric calculation over the entire shock
  surface -- where $P_\mathrm{c}$ is the CR pressure and $t_\mathrm{sn}$ is
the SNR age \citep[see][]{bv04b}.  The value of the magnetic field at the
current epoch $B_0(t_\mathrm{sn})$ is determined as a result of the best fit of
the measured synchrotron spectrum (see below).  In the downstream region the
magnetic field is assumed to be frozen into the gas  
with the postshock field value $B_2 \approx \sigma
B_0$, where $\sigma$ is the shock compression ratio. Under these assumptions
the magnetic field strength varies only slightly within the downstream
region of swept-up ISM. Therefore below the value $B_\mathrm{d}=B_2$ is
used as the interior magnetic field strength.  Also the notations
$B_\mathrm{d}$ and $B_0$ are used for their values at the current SNR epoch
$t=t_\mathrm{sn}$.

Since the properties of accelerated CR nuclear and electron spectra and their
dependence on the relevent physical parameters, as well as the dynamical
properties of the system were described in detail in previous papers
\citep[e.g][]{bkv02,kbv05}, they will not be discussed here (see however the
Appendix for a summary).

%
\begin{figure}
 \includegraphics[width=.47\textwidth]{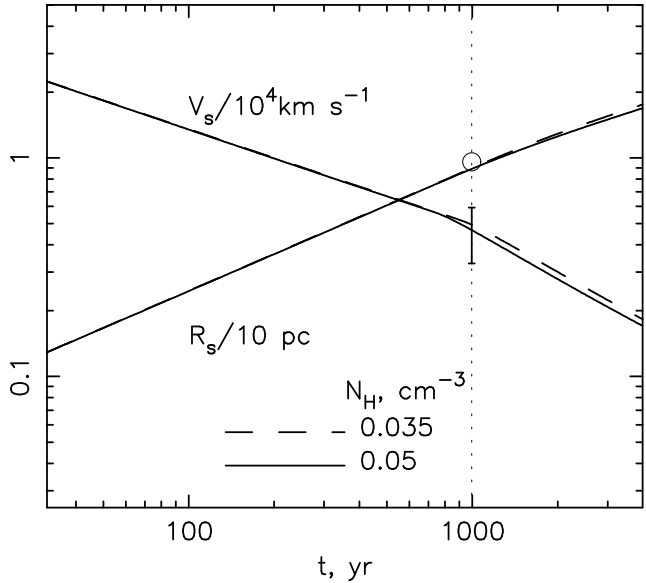}
 \caption{Gas dynamical evolution for the two values of the total explosion
   energy $E_\mathrm{sn}= 1.8\times 10^{51}$~erg and $E_\mathrm{sn}= 1.5\times
   10^{51}$ that correspond to the values $N_\mathrm{H}=0.05~\mbox{cm}^{-3}$
   and $N_\mathrm{H}=0.035~\mbox{cm}^{-3}$ of the ISM hydrogen number density,
   respectively. The quantities $R_\mathrm{s}$ and $V_\mathrm{s}$ denote the
   shock radius and the shock velocity, their time evolution being fitted to the
   observed values shown. The vertical dotted line indicates the present epoch.}
\label{fig1}
\end{figure}
\noindent The values of the SN explosion energy $E_\mathrm{sn}= 1.8\times
10^{51}$~erg and $E_\mathrm{sn}= 1.5\times 10^{51}$~erg are taken to fit the
observed shock size $R_\mathrm{s}$ and shock speed $V_\mathrm{s}$
\citep{moffett93} at the current epoch $t\approx 10^3$~yr (see Fig.1) for the
ISM hydrogen number densities $N_\mathrm{H}=0.05~\mbox{cm}^{-3}$ and
$N_\mathrm{H}=0.035~\mbox{cm}^{-3}$, respectively. These densities are
consistent with the observed level of the VHE emission (see below).  Note that
the calculation presented in Fig.1 corresponds to the best fit value of the
amplified upstream magnetic field strength $B_0=30$~$\mu$G, even though
the shock size $R_\mathrm{s}$ and shock speed $V_\mathrm{s}$ are quite
insensitive to $B_0$. The resulting current total shock compression ratios
$\sigma$ for $N_\mathrm{H}=0.05~\mbox{cm}^{-3}$ and
$N_\mathrm{H}=0.035~\mbox{cm}^{-3}$are $\sigma = 4.9$~and 4.7, respectively,
whereas the subshock compression ratios $\sigma_\mathrm{s}$~ are both close to
$\sigma_\mathrm{s}=3.7$.

As the most reliable estimate for the distance the
  value $d=2.2$~kpc was taken \citep{winkler03}.

  All these results stem of course mainly from the gas dynamics part of the
  governing equations.

%
\begin{figure}
 \includegraphics[width=.47\textwidth]{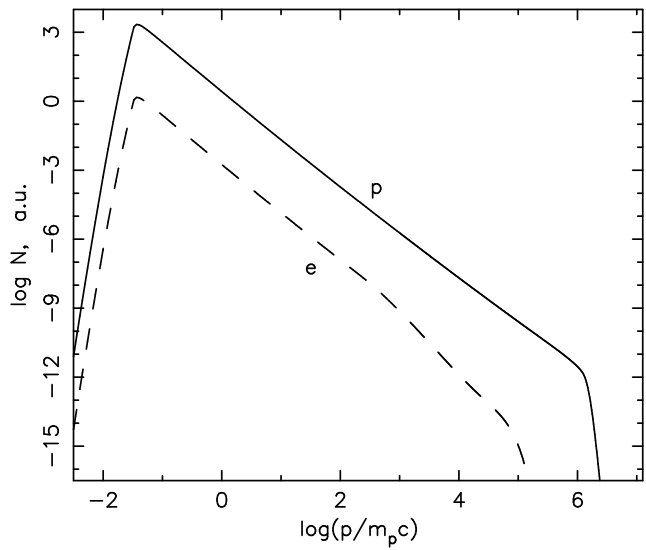}
 \caption{Overall (=spatially integrated) spectra of energetic particles. Solid
   and dashed lines correspond to protons (p) and electrons (e), respectively.}
\label{fig2}
\end{figure}

 Fig.2 shows the volume-integrated momentumg distributions of
energetic protons (representative for the nuclear component) and electrons. The
proton cutoff 
energy $\epsilon_\mathrm{max}^\mathrm{p}$
is above $10^{15}$~eV. The electron spectral index $\gamma \approx
2$ equals that of the protons below the onset of synchrotron losses at the
energy $\epsilon_l \approx 1$~TeV. The electron cutoff energy 
$\epsilon_\mathrm{max}^\mathrm{e}$ is near
$10^{14}$~eV.

%
\begin{figure}
 \includegraphics[width=.47\textwidth]{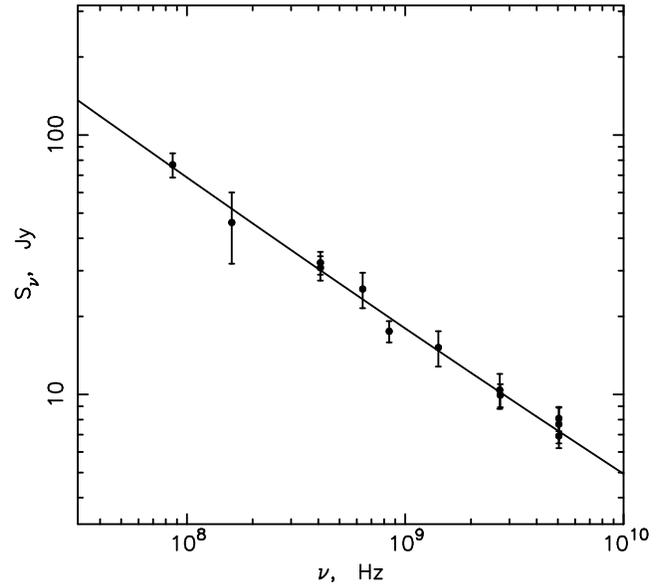}
 \caption{Differential synchrotron radio emission flux as a function of
   frequency, calculated for the ISM hydrogen number density
   $N_\mathrm{H}=0.05~\mbox{cm}^{-3}$ at the proton injection rate $\eta =
   3\times 10^{-4}$, electron to proton ratio, $K_\mathrm{ep} = 4\times
   10^{-4}$, and magnetic field strength $B_0=30$~$\mu$G upstream of the
   subshock . The observed radio emission flux from a compilation of
   \citet{allen04} \citep[see also][]{allen01,allen08} is shown.}
\label{fig3}
\end{figure}

\begin{figure*}
\center
 \includegraphics[width=.95\textwidth]{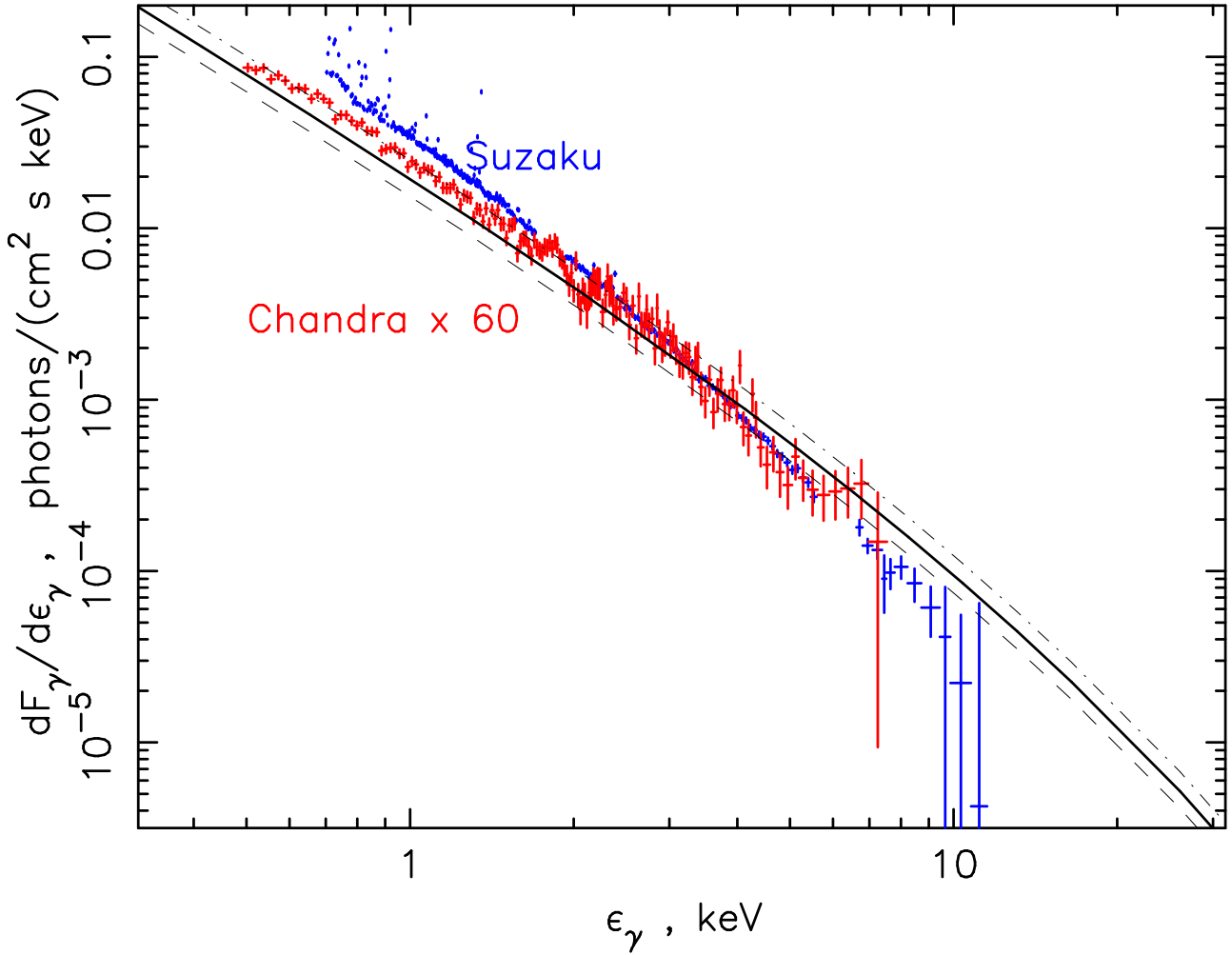}
 \caption{X-ray synchrotron emission flux as a function of photon energy for
   the same case as in Fig.1 (solid line).  Calculations for $B_0=27$~$\mu$G
   ({\it dashed line}) and for $B_0=33$~$\mu$G ({\it dash-dotted line}) are
   also presented.  The observed X-ray flux from Chandra {\it in red
       color}  for a small region of the bright northeastern
   rim of SN~1006 \citep{allen04}, and the global X-ray spectrum {\it in
       blue color} as observed by Suzaku \citep{bamba08} are shown. The
   Chandra X-ray flux was normalized such as to be consistent with the Suzaku
   flux for energies $\epsilon > 2$~keV.}
\label{fig4}
\end{figure*}

\begin{figure}[t]
 \includegraphics[width=.47\textwidth]{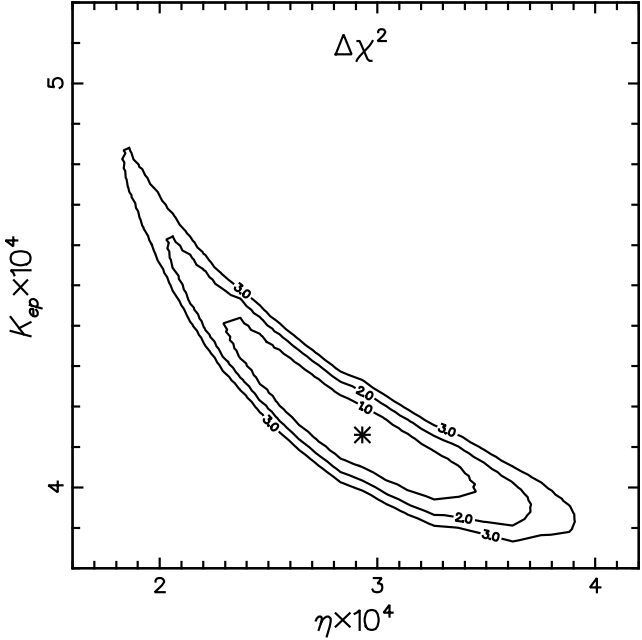}
 \caption{Contours of equal deviations $\Delta \chi^2$
(shown on the curves) from the minimum value of $\chi^2$, in the plane 
     that is spanned by the electron to proton ratio $K_\mathrm{ep}$ 
 and the proton injection rate $\eta$. The {\it asterisk} corresponds
   to the best fit values $\eta=3\times 10^{-4}$ and $K_\mathrm{ep}= 4\times
   10^{-4}$ with the minimum value $\chi^2/\mathrm{dof} = 1.3$. The fit
     uses jointly the radio and nonthermal X-ray data.}
\label{fig5}
\end{figure}

\noindent Figs.~\ref{fig3},~\ref{fig4} and ~\ref{fig5} illustrate the
consistency of the synchrotron spectrum, calculated for the ISM number density
$N_\mathrm{H}=0.05~\mbox{cm}^{-3}$ with the best set of parameters ($\eta =
3\times 10^{-4}$, $K_\mathrm{ep} = 4\times 10^{-4}$, $B_0=30$~$\mu$G), with the
observed spatially integrated spectra. Note that we use the differential
  synchrotron fluxes in the radio ($S_{\nu}$) and in the X-ray
  ($dF_{\gamma}/d\epsilon_{\gamma}$) range in a slightly different form. They
  are however simply connected by the relation $\nu
  S_{\nu}=\epsilon_{\gamma}^2dF_{\gamma}/d\epsilon_{\gamma}$.

  It should be noted that the values $\alpha>0.5$ of the differential radio
  spectral index $\alpha= -d\ln S_{\nu}/d\ln \nu$, as observed in young SNRs,
  require a curved electron spectrum that hardens to higher energies as
  predicted by nonlinear shock acceleration models, as originally emphasized by
  \citet{reynolds92}. To have $\alpha=0.57$ in the radio range, as observed for
  SN~1006, one needs efficient CR acceleration with a proton injection rate
  $\eta=3\times 10^{-4}$ which leads to the required shock modification, and a
  high interior magnetic field $B_\mathrm{d}\ge 130$~$\mu$G is required
  \citep{bkv02,vbk05,kbv05}. The hardening of the observed radio synchrotron
  spectrum for SN~1006 has been recently demonstrated by \citet{allen08}.
  Detailed X-ray synchrotron spectral measurements are, however, required to
  find the optimum value of the magnetic field strength $B_\mathrm{d}$
  \citep{kbv05} since for a given fit of the calculated synchrotron spectrum in
  the radio range the X-ray synchrotron amplitude is very sensitive to
  $B_\mathrm{d}$.  The reason is the following: in the polar regions, where the
  scattering mean free path decreases down to the particle gyroradius (Bohm
  limit), the spectrum of accelerated electrons at the shock front is quite
  insensitive to the magnetic field strength $B_\mathrm{d}$ if all other
  relevant parameters are fixed; the exception is their maximal energy
  $\epsilon_\mathrm{max}^\mathrm{e}\propto V_\mathrm{s}B_\mathrm{d}^{-1/2}$
  which is due to synchrotron losses \citep[e.g.][]{bv04b}. Since all energy of
  the electrons at high energies $\epsilon>\epsilon_\mathrm{l}\propto
  B_\mathrm{d}^{-2}$, where synchrotron losses are significant, is rapidly and
  completely transformed into the synchrotron emission, the same is true for
  the high frequency ($\nu>\nu_\mathrm{l}\propto
  \epsilon_\mathrm{l}^2/B_\mathrm{d}$) part of the synchrotron spectrum
  including its cutoff part, because the cutoff frequency
  $\nu_\mathrm{max}\propto \epsilon_\mathrm{max}^\mathrm{e}B_\mathrm{d}^{1/2}$
  does not depend on $B_\mathrm{d}$.  (In the case $B_0=30$~$\mu$G and
  $B_\mathrm{d}=150$~$\mu$G we have $\epsilon_\mathrm{max}^\mathrm{e}\approx
  10^{14}$~eV and $\nu_\mathrm{max}\approx 10^{18}$~Hz.)  However, varying the
  magnetic field strength $B_\mathrm{d}$ changes the amplitude of the overall
  calculated synchrotron spectrum at high frequencies $\nu>\nu_\mathrm{l}$ in
  order to keep the fit of the measured radio spectrum at low frequencies.
  Therefore the high frequency part of the spectrum $S_{\nu}$ changes due to
  the variation of the breaking frequency $\nu_\mathrm{l}\propto
  B_\mathrm{d}^{-3}$. Since below $\nu_\mathrm{l}$ the synchrotron spectrum has
  the form $S_{\nu}\propto \nu^{-\alpha}$ with $\alpha\approx 0.5$ we conclude
  that the high frequency part of the synchrotron spectrum depends on
  $B_\mathrm{d}$ like $S_{\nu}(\nu_\mathrm{l})\propto B_\mathrm{d}^{-3/2}$. The
  amplitude $S_{\nu}(\nu_\mathrm{l})$ is therefore the main determinant of the
  value of $B_\mathrm{d}$. Together with the observed synchrotron amplitude in
  the radio frequency region, where radiative losses play no role yet, also the
  amplitude of the nonthermal electron distribution and thus the
  electron/proton ratio ratio $K_\mathrm{ep}$ is then determined.  Note that
  the synchrotron spectrum is almost insensitive to the ISM density
  $N_\mathrm{H}$ \citep[e.g.][]{kbv05}.
%

To obtain the fit to the data shown in Fig.~\ref{fig3} and Fig.~\ref{fig4} it
is asserted that only the highest-energy part of the observed global X-ray
spectrum, corresponding to $\epsilon > 2$~keV, is of predominantly nonthermal
origin. Towards lower energies $\epsilon < 2$~keV the contribution of thermal
emission to this X-ray spectrum progressively increases, as clearly indicated
by the Suzaku spectrum.  For these considerations the X-ray Chandra flux from a
small region of the bright north eastern rim of SN~1006 \citep{allen04} is
used, where the contributions from thermal X-rays are presumably minimal
\citep{allen04, cassam08}. In order to apply it to the whole remnant, this
X-ray flux was normalized to be consistent at energies $\epsilon > 2$~keV with
the global X-ray spectrum as observed by Suzaku \citep{bamba08}. This
consistency in fact exists.  For quantitative purposes the normalised Chandra
spectrum above 2~keV is used, under the assumption that it corresponds to pure
nonthermal emission.  Comparison with the experimental X-ray data shows in
Figs.~\ref{fig3} and \ref{fig4} that the optimum magnetic field value is
$B_0=30$~$\mu$G, and corresponds to a downstream field $B_{\mathrm d}\approx
150$~$\mu$G. This is in agreement with the field amplification that is implied
by the filamentary structures in hard X-rays \citep{vbk05}.

Fig.~\ref{fig5} shows the contours of $\chi^2$ characterizing the quality of
fit of the radio spectrum, and of the X-ray spectrum for $\epsilon > 2$~keV,
for different values of $\eta$ and $K_\mathrm{ep}$. One can see that the
quality of the fit is rather good, allowing only a rather small range of these
parameters.

The quality of fit of the radio spectrum, and of the X-ray spectra for
$\epsilon > 2$~keV, for $B_0=30$~$\mu$G is characterised by the value
$\chi^2/\mathrm{dof} =1.3$. An increase $\Delta \chi^2 \approx 1$ of $\chi^2$
(see Fig.~\ref{fig4}) implies a change of the magnetic field strength value by
only 10 \%.

The \gr morphology, as found in the H.E.S.S. measurements
  \citep{naumann09}, is consistent with the prediction of a polar cap geometry
  by \citet{vbk03}. Such a geometry has been also found experimentally from an
  analysis of the synchrotron morphology in hard X-rays by \citet{ro04} and
  \citet{cassam08}. This means that the \gr emission calculated in the
  spherically symmetric model must be renormalized (reduced) by a factor
  $f_\mathrm{re} \approx 0.2$, as in \citet{kbv05}. The
renormalization factor is applied here.

This morphology is also the key argument for the existence of
an energetically dominant nuclear CR component in SN~1006, because only such a
component can amplify the magnetic field to the observed degree. If on the
contrary the existing energetic electron component would have to drive the
magnetic field amplification all by itself then we would, as a minimum, require
$P_\mathrm{e}\gg B_0^2/(8\pi)$, where $P_\mathrm{e}$ is the pressure of CR
electrons at the shock front.  Clearly $P_\mathrm{e}\lsim K_\mathrm{ep}
P_\mathrm{p}$, where $P_\mathrm{p} \approx P_\mathrm{c}$ is the pressure of CR
protons and $P_\mathrm{c}$ is the pressure of the nuclear component. Since
$P_\mathrm{c}\approx 0.5\rho_0V_\mathrm{s}^2$ we have $P_\mathrm{e}\approx 0.07
B_0^2/(8\pi)$ for $N_\mathrm{H}=0.05~\mbox{cm}^{-3}$ and $V_\mathrm{s} \approx
4000$~km/s.  Therefore accelerated electrons are not able to amplify the
magnetic field to the required level by a large margin. On the other hand, the
effectively accelerated nuclear component has the pressure $P_\mathrm{c}\approx
200 B_0^2/(8\pi)$ and can therefore readily amplify the field from a purely
energetic point of view.

The amplified field must also be the reason for the corresponding polar
cap-type morphology of the synchrotron emission in hard X-rays. Even though the
weak radio synchrotron emission, essentially all around the periphery of
SN~1006 \citep{ro04,cassam08}, demonstrates that electrons are at least to some
extent injected everywhere into the acceleration process over the shock, they
reach the multi-TeV energies for X-ray synchrotron emission only in the polar
caps, where the amplified field allows their acceleration to these energies. In
the equatorial region the upstream magnetic field strength corresponds to its
ISM value, at the height of SN~1006 above the Galactic plane probably not more
than $3 \mu$G. Particle scattering is comparatively weak and the maximum
  electron energy is expected to lie below the limit, where the acceleration
  rate equals the synchrotron loss rate. Then the maximum electron energy is --
  like that of the nuclear particles -- determined by the system geometry
  \citep{ber96}; an explicit calculation can be found in \citet{bv04b}.
Assuming then for simplicity that the spatial diffusion coefficient 
  has the form $\kappa = \delta \cdot \kappa_\mathrm{Bohm}$, with a constant
factor $\delta > 1$, requires $\delta > 4$ even for this low field
  strength in order to reduce the equatorial maximum synchrotron frequency
$\nu_\mathrm{max}$ to $\nu_\mathrm{max} < 10^{17}$~Hz, for an azimuthally
uniform shock velocity $V_\mathrm{s} \approx 4000$~km/s. Such a plausible
reduction of $\nu_\mathrm{max}$ is consistent with the findings of
\citet{ro04}, see also \citet{miceli09}.

The only important parameter which can not be determined from the analysis of
the synchrotron emission data is the external density $N_\mathrm{H}$.
Therefore we have performed the calculations for the pair of values
$N_\mathrm{H}=0.05~\mbox{cm}^{-3}$ and $N_\mathrm{H}=0.035~\mbox{cm}^{-3}$ which
seems to bracket the density range consistent with the H.E.S.S. \gr
measurements.

\begin{figure}[t]
 \includegraphics[width=.47\textwidth]{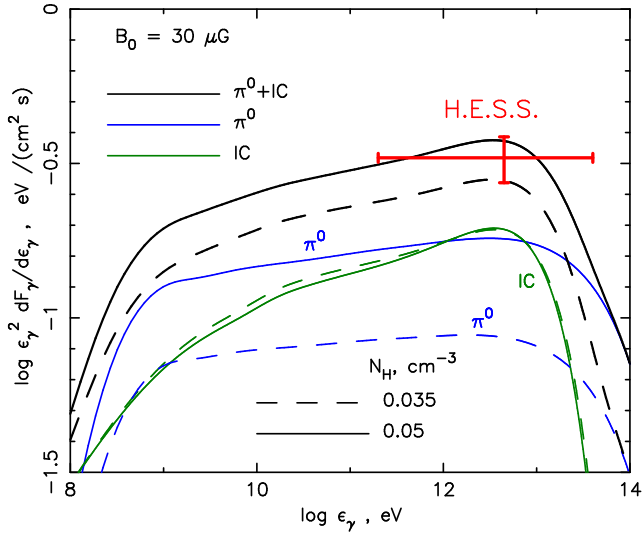}
 \caption{Total ($\pi^0$-decay + IC) ({\it black lines}), $\pi^0$-decay ({\it
     blue lines}), and IC ({\it green lines}) differential $\gamma$-ray energy
   fluxes as a function of $\gamma$-ray energy, calculated for ISM hydrogen
   number densities $N_\mathrm{H}=0.05~\mbox{cm}^{-3}$ ({\it solid lines}) and
   $N_\mathrm{H}=0.035~\mbox{cm}^{-3}$ ({\it dashed lines}) for the parameters
   $\eta=2.9\times 10^{-4}$, $K_{ep}= 4.1\times 10^{-4}$ and $B_d=150$~$\mu$G,
   derived from the fit of the synchrotron spectrum. The H.E.S.S.  value
   \citep{naumann09} is shown as well. }
\label{fig6}
\end{figure}

Fig.~\ref{fig6} shows the total ($\pi^0$-decay plus inverse Compton (IC)),
  and seperately the $\pi^0$-decay and the IC $\gamma$-ray energy spectra of
the remnant, calculated for $N_\mathrm{H}=0.05~\mbox{cm}^{-3}$ and
$N_\mathrm{H}=0.035~\mbox{cm}^{-3}$.

H.E.S.S. has reported the measured value of the energy flux $\Phi (\epsilon_1,
\epsilon_2)=2.5\times 10^{-12}$~erg/(cm$^2$~s) of \grs with energies
$\epsilon_1 \le \epsilon_{\gamma}\le \epsilon_2$, where $\epsilon_1 =
  0.2$~TeV and $\epsilon_2 = 40$~TeV. It is related with the differential \gr
energy spectrum according to the expression
\[
\Phi(\epsilon_1, \epsilon_2)=
\int_{\epsilon_1}^{ \epsilon_2}
 \epsilon_{\gamma}^2 
\frac{dF_{\gamma}}{d\epsilon_{\gamma}}d\ln \epsilon_{\gamma}
=\ln 
\frac{\epsilon_2 }{\epsilon_1}
<\epsilon_{\gamma}^2 \frac{dF_{\gamma}}{d\epsilon_{\gamma}}>.
\] 

In a selfconsistent picture the
  theoretically predicted form of the differential spectrum
  $dF_{\gamma}/d\epsilon_{\gamma}$, cf.  Fig. 6 is used to calculate
the mean experimental value of the differential energy flux
$<\epsilon_{\gamma}^2 dF_{\gamma}/{d\epsilon_{\gamma}}>= \Phi(\epsilon_1,
\epsilon_2)/\ln(\epsilon_2/\epsilon_1)$, shown in Fig.~\ref{fig6}.

Note that the peak in the $\gamma$-ray energy spectrum at $\epsilon_{\gamma}
\approx 5$~TeV and the quasi-exponential cutoff at higher energies is due to
the substantial contribution of the IC component: at TeV energies the IC
component contributes about half of the total \gr flux for
$N_\mathrm{H}=0.05~\mbox{cm}^{-3}$, whereas for
$N_\mathrm{H}=0.035~\mbox{cm}^{-3}$ it is already two thirds.  Since the
maximal proton energy, reached at some time during the evolution, is about
$\epsilon_\mathrm{max}\approx 2\times 10^{15}$~eV, the corresponding
$\pi^0$-decay component has a power law form
$dF_{\gamma}/d\epsilon_{\gamma}\propto \epsilon_{\gamma}^{-\gamma}$ with
$\gamma\sim 2$ up to the cutoff energy $\epsilon_{\gamma}\approx
10^{14}$~eV. The cutoff energy is defined here as that energy, where the
spectral energy density has dropped by a factor $1/e$ from its maximum value
which given by log $\epsilon_{\gamma}^2 dF_{\gamma}/{d\epsilon_{\gamma}}
\approx -0.8$.

According to Fig.~\ref{fig6} the H.E.S.S. data are consistent with an ISM
number density from quite a narrow interval $0.035\le N_\mathrm{H}\le
0.05~\mbox{cm}^{-3}$, since for the theoretically derived \gr spectrum we have
$\Phi=2.1\times 10^{-12}$~erg/(cm$^2$~s) and $\Phi=2.9\times
10^{-12}$~erg/(cm$^2$~s) for $N_\mathrm{H}=0.035~\mbox{cm}^{-3}$ and
$N_\mathrm{H}=0.05~\mbox{cm}^{-3}$ respectively. It should be noted that the
corresponding explosion energy $E_\mathrm{sn}\approx 1.7\times 10^{51}$~erg is
close to the upper end of the typical range of type Ia SN explosion energies
that vary by a factor of about two \citep{gamezo05,brs06}.

\citet{acero07} find the value $N_\mathrm{H}\approx 0.05~\mbox{cm}^{-3}$ on the
basis of X-ray measurements. The above interval is consistent with their
result.

From Fig. 6 the \gr spectrum produced by the nuclear CRs is very close to the
IC emission spectrum produced by CR electrons alone. Since the differential
energy spectrum of freshly accelerated nuclear particles and electrons is
rather close to a spectrum $N_\mathrm{e}\propto \epsilon^{-2}$, and since the
electrons with energies $\epsilon>\epsilon_\mathrm{l}\approx 1$~TeV
subsequently undergo significant synchrotron cooling in the interior, leading
to the spectrum $N_\mathrm{e}\propto \epsilon^{-3}$, not only the amplitude but
also the shape of these two components are very similar within the energy
interval $10^{10}<\epsilon_{\gamma}<10^{13}$~eV. Therefore the VHE \gr spectrum
alone is not able to discriminate between the hadronic $\pi^0$-decay and the
leptonic IC \gr components. However, it was already shown by \citet{kbv05} that
such a low VHE emission flux, with a highly depressed IC \gr flux, is only
possible if the nuclear CR component is efficiently produced with accompanying
strong magnetic field amplification. In the framework of the interpretation
developed in this paper the most direct evidence for the energetic dominance of
a nuclear energetic particle component is the observed \gr morphology. It
corresponds to the theoretical prediction and is consistent with all other
measurements.

A last point concerns the radial extent $R_\mathrm{c}$ of the contact
discontinuity between ejected and swept-up mass relative to the radius
$R_\mathrm{s}$ of the SNR blast wave, cf. recent data presented and discussed
by \citet{cassam08}. The ratio $R_\mathrm{s}/R_\mathrm{c}$ is given as $1.04
\pm 0.03$ outside the synchrotron rims, and essentially as $1$ in the region
within the synchrotron rims.  In the case of Tycho's SNR we have discussed in
quantitative detail \citep [e.g.][]{vbk08} the reduction of this ratio compared
to a pure gas shock as a result of the considerable shock modification produced
by accelerated nuclear CRs, which leads to the increase of the shock
compression. Qualitatively such considerations agree with the experiment for
SN~1006: the ratio $R_\mathrm{s}/R_\mathrm{c}$ is larger in the equatorial
region, where CR injection/acceleration is inefficient, and it is essentially
smaller within the polar region, where CRs are efficiently produced.  While for
Tycho's SNR particle acceleration gives a good theoretical explanation of the
relatively small ratio within our model, this is clearly impossible
quantitatively for the above numbers in the case of SN 1006. \citet{cassam08}
believe that the value of the contact discontinuity radius $R_\mathrm{c}$ was
somehow overestimated "... since our measurements are likely to be affected by
projection and other effects, ...". On the other hand, very recently
\citet{miceli09} have found numbers $R_\mathrm{s}/R_\mathrm{c}$ of the order of
1.1 which are in the expected range. Future work will have to resolve this
difference.

\section{Summary}

Since the relevant astronomical parameters as well as the synchrotron spectrum
of SN~1006 are measured in impressive detail it is possible to determine the
values of the relevant physical parameters with the appropriate accuracy for
this SNR: proton injection rate $\eta=(2.9\pm 0.6)\times 10^{-4}$, electron to
proton ratio $K_{ep}= (4.1\pm 0.3)\times 10^{-4}$ and downstream magnetic field
strength $B_d=(150\pm 15)$~$\mu$G.

As a result the flux of TeV emission detected by H.E.S.S. is consistent with
the ISM number density $0.035\le N_\mathrm{H}\le 0.05~\mbox{cm}^{-3}$. The
corresponding hydrodynamic SN explosion energy $E_\mathrm{sn}= 1.7\times
10^{51}$~erg is close to the upper end $E_\mathrm{sn}= 1.6 \times 10^{51}$~erg
of the typical range of type Ia SN explosion energies that vary by a factor of
about two. Also the magnetic field amplification properties of this SNR are
well understandable as the result of azimuthal variations of ion injection over
the projected SNR circumference and corresponding acceleration which lead to a
polar cap-type morphology for the X-ray synchrotron as well as the \gr
emission. As a consequence, the recent H.E.S.S. measurement of a dipolar
  morphology also of the TeV \gr emission is considered as the most significant
  empirical evidence for the existence of an energetically dominant nonthermal,
  nuclear component.  

In conclusion SN~1006 appears to represent the first example where a high
efficiency of nuclear CR production, required for the Galactic CR sources, is
consistently established.



\begin{acknowledgements} 
  This work has been supported in part by the Russian Foundation for Basic
  Research (grants 06-02-96008, 07-02-0221). The authors are grateful to
  Drs. Aya Bamba and Glen Allen who provided them with the most recent X-ray
  data from the Chandra and Suzaku satellite observatories. The authors also
  thank Dr. Vladimir Zirakashvili for discussions regarding
  electron acceleration in SN 1006. EGB acknowledges the hospitality of the
  Max-Planck-Institut f\"ur Kernphysik, where part of this work was carried
  out.
\end{acknowledgements}

\appendix
\section{ }

The relation between the model calculations in spherical symmetry (1D) and the
physical characteristics of the system in the presence of a symmetry-breaking
magnetic field in the circumstellar medium have been discussed earlier
\citep{vbk03}. In order to clarify the necessary corrections to the
1D-calculation for SN~1006 a qualitative discussion of the physical situation
is given here.

A key feature is the empirical fact that the force density of the magnetic
field in the diffusive shock acceleration process can be neglected compared to
the gas ram pressure, the thermal pressure and the CR pressure. The average
magnetic field configuration can thus in principle be calculated purely
kinematically from the induction equation. However, the magnetic field
direction influences the injection at least of heavy ions in a significant
way. In a quasi-parallel shock, where the angle $\Theta_\mathrm{nB}$ between
the locally mean field and the shock normal is small compared to $90^{\circ}$,
the required velocity of suprathermal ions from the downstream region, required
to be able to cross the shock into the upstream region, is much smaller than in
a quasi-perpendicular shock, where $\Theta_\mathrm{nB}$ is close to
$90^{\circ}$. For a steeply falling velocity distribution of the shocked
downstream gas this leads to effective suppression of ion injection in
quasi-perpendicular shocks \citep{ellison95,malkov95}. The acceleration of
injected particles, on the other hand is efficient for all $\Theta_\mathrm{nB}$
for non-relativistic shocks, as long as $\cos \Theta_\mathrm{nB} \gg
V_\mathrm{s}/c$, where $V_\mathrm{s}$ and $c$ denote the shock velocity and the
speed of light, respectively \citep{drury83}. The injection of electrons is not
yet well understood. From the observed azimuthal distribution of the radio
synchrotron emission in SN~1006 \citep[e.g.][]{cassam08} electron injection
seems to operate for all angles $\Theta_\mathrm{nb}$, although apparently less
efficiently in the quasi-perpendicular region. This is the least restrictive
assumption and it is made here in the sequel.

A second point is that in the instantaneously quasi-parallel part of the
magnetic flux tubes, in which ions are injected, the accelerated particles
drive a current against the flow of the upstream gas that generates strong
magnetic fluctuations and amplifies the magnetic field, as referred to in the
main text. This enhances the acceleration and leads to a pressure
$P_\mathrm{c}$ of the energetic nuclear particles that is comparable to the gas
ram pressure for strong shocks, e.g. in young SNRs, as well as to heating of
the gas by wave dissipation, both of the non-resonant and of the resonant
Alfv\'enic modes \citep{mv82,lb00,
  belll01,bell04,pelletier06,zirakashvili08}. Especially the formation of
density fluctuations and corresponding secondary shocks is viewed here as a
reason for strong wave dissipation \citep{ber08,vbk08}. The pressure of
accelerated electrons is negligible in comparison. This implies a strong
hadronic \gr emission from that part of these flux tubes, which became at some
time quasi-parallel. This hadronic flux is negligible elsewhere because there
are few energetic nuclei there. A certain amount of
cross-field diffusion will, however, also populate neighboring flux tubes and
somewhat smear out the acceleration boundaries \citep{vbk03}. Accelerated
electrons will emit strong synchrotron radiation in the enhanced magnetic field
regions, but may also yield a more moderate synchrotron flux elsewhere. The
reason is that by assumption they are injected and accelerated to some extent
``everywhere'' on the shock surface, albeit to considerably lower maximum
energies in the quasi-perpendicular regions. This is the result of the lack of
self-excited magnetic field fluctuations and of field amplification (see
section 2).

The third point regards differences in overall dynamics of the quasi-parallel
and quasi-perpendicular regions. Conservation of the overall mass,
momentum and energy fluxes allows only a different {\it partition} within the
momentum and energy fluxes between their kinetic, thermal and nonthermal
components, but the sums of these fluxes always remain the same. However, the
compressibility of the relativistic CR ``gas'' is higher than that of the
nonrelativistic thermal gas, and therefore this different partition implies
dynamical differences between the quasi-parallel and the quasi-perpendicular
regions.

When primarily the cumulative effect of the higher compressibility counts, like
in the relative distance between the outer SNR blast wave and the following
contact discontinuity, then the different partitions in a CR-modified shock and
a pure gas shock play the determining role.

However, in the present context, where the main characteristics of the
radiation effects and of their spatial distributions are to be evaluated, such
dynamical differences can to first approximation be neglected. In practice, for
SN 1006, the overall shock compression ratio changes from $\sigma = 4$ without
CR production to $\sigma \sim 5$ with efficient CR production, given the
  assumed strong gas heating in the shock precursor.

Thus, for the evaluation of the hadronic \gr emission, only those flux tubes
from the 1D-configuration gas dynamic configuration are considered which
connect to the part of the present phase shock surface, where ion injection is
efficient. This requires the correction factor $f_\mathrm{re}<1$. Since the
particle transport along magnetic field lines is primarily diffusive, adiabatic
expansion of the thermal gas leads to adiabatic losses of the energetic
particles and weakens the interior particle distribution. In addition
  almost all flux tubes have been originally quasi-perpendicular. The
  corresponding parts of these flux tubes are now in the deeper interior of the
  SNR and therefore, to first approximation, lack there accelerated nuclei.
For electrons also the radiative losses increase towards the interior where the
`older' particles reside. These arguments show that both the synchrotron and
the \gr emission are concentrated near the shock surface rather than in the
interior. As long as the ejected mass is not negligible compared to the
swept-up mass, the ejecta volume in the interior with its extremely small
magnetic field strength will appear as a region with particularly little
radiation. For SN~1006 with its supposedly simple magnetic field geometry, the evaluation of the ion injection fraction
$f_\mathrm{re}$ of the shock surface has been given by \citet{vbk03}.

Also the electron spectrum is calculated in spherical symmetry from Eq.3
  of \citet{bkv02}. However the actual evolutionary phase of SN 1006 appears to
  be still rather in the transition from the sweep-up phase to what one can
  losely call a Sedov phase. Being pushed away by the ejecta the density of the
  shocked circumstellar medium has remained rather uniform. In the
  quasi-parallel flux tubes, downstream of the shock, with massively injected
  nuclear particles, the field $B_\mathrm{d}$  should therefore
  be strong and roughly uniform if field dissipation is negligible.  In the
  quasi-perpendicular regions on the other hand, the field is only about four
  times larger than in the ambient circumstellar medium. Finally, in the ejecta
  mass the magnetic field strength is presumably very small at the present
  time. Taken all this together, the overall magnetic field strength in the SNR
  is far from uniform. Yet what counts primarily for the integrated synchrotron
  emission are the quasi-parallel regions.

It had indeed been assumed in past work \citep{bkv02,kbv05} that to first
approximation only those magnetic flux tubes play a role in the spatially
integrated emission which end at the forward shock in the polar caps. Therefore
this integral already contains the correction factor $f_\mathrm{re}$
implicitely and can be directly compared to the observed integral flux. The key
consideration in all of this is that the SNR consists of a strongly CR-modified
part with strong magnetic field amplification and a part which is essentially
unmodified. The correction factor changes only the amplitude but not the {\it
  form} of the calculated spectrum.

Adding a small population of electrons accelerated in the unmodified shock of
the quasi-perpendicular regions would smear out the specifically nonlinear
properties of the synchrotron spectrum. Therefore, the observed synchrotron
spectrum exhibits the minimum nonlinear modifications which in reality should
be even somewhat stronger. This means that the derived ion injection rate,
magnetic field amplification and electron/proton ratio are lower limits for the
quasi-parallel regions.



\end{document}